 \documentclass[final,3p,times,twocolumn]{elsarticle}

\usepackage{amsmath}
\usepackage{amssymb,amsfonts,textcomp}

\biboptions{compress}

\makeatletter
\def\ps@pprintTitle{%
  \let\@oddhead\@empty
  \let\@evenhead\@empty
  \let\@oddfoot\@empty
  \let\@evenfoot\@oddfoot
}
\makeatother

\begin{document}

\begin{frontmatter}



\title{Grain boundary segregation of C, N and O in hcp titanium from first-principles}


\author{D.A. Aksyonov}
\author{A.G. Lipnitskii}
\author{Yu.R. Kolobov}

\address{The Center of nanostructured materials and nanotechnologies, Belgorod state university, Belgorod, Russian Federation}

\begin{abstract}

It is believed that grain boundary segregation of light interstitials 
can serve as the possible mechanism of thermal stability in 
commercially pure nanostructured titanium alloys. In this paper, using first-principles calculations, 
we show that independent segregation of C, N and O atoms at $\Sigma 7$ 
high angle grain boundary in $\alpha$-Ti is energetically unfavourable. The presence of interstitial 
elements near the grain boundary plane
results in the increase of the grain boundary width and specific formation energy. 
\end{abstract}


\end{frontmatter}


\section{Introduction}
\label{intro}

An increased (decreased) concentration of impurities at grain boundaries (GB), known as segregations
can considerably influence various properties of 
polycrystalline materials \cite{Lejcek2010}. Besides causing an embrittlement effect, they can increase grain boundary cohesion and 
significantly improve thermal stability of nanocrystalline materials 
\cite{Weissmuller1992, Weissmuller1993, Farber2000,Liu2001,Choi2005,Detor2007, VanLeeuwen2010, Pellicer2011, Chookajorn2012,Tang2012, Atwater2013}. 
Hence, it is important for the design and optimisation 
of nanomaterials, which show a large specific area of interfaces that can give rise to segregations, to understand the influence of segregations 
on materials properties.

Commercially pure nanostructured $\alpha$-titanium (CP-nTi) alloys with hcp lattice structure have a high biocompatibility and superior strength 
 in comparison to coarse-grained Ti \cite{Handtrack2008, Ivanov2011a}. Being a promising material for medical lifetime dental implants the problem of thermal 
stability is very important for CP-nTi. There are no data in literature about long term behaviour of this material, because the mechanisms that 
are thermally stabilising nTi are not understood yet. Due to a technologically necessary heat treatment of implants during production, an 
improvement of their microstructure stability is a significant task, which needs first of all a detailed understanding of existing mechanisms of 
grain boundary stabilisation. CP titanium alloys (Grade 1-4) contain a sufficiently large amount of light element impurities such as C, N and O 
\cite{Aksyonov2012}. Hence, two most probable mechanisms of thermal stability are possible in nTi: (i) grain boundary pinning by precipitates 
(dispersed particles of carbides, nitrides, etc. formed from solid solution) and 
(ii) lowering of driving force for grain growth and reduction of grain boundary mobility due to segregation of 
existing impurities \cite{Weissmuller1992}. 
The possibility of formation of Ti-C precipitates in $\alpha$-Ti was considered in our recent work \cite{Aksyonov2012}. The 
purpose of the current work is to continue the investigation of thermal stability mechanisms in CP-nTi, 
considering the case of light impurity
segregation at $\alpha$-Ti grain boundaries in comparison with previous results of Ti-C particles formation.

Semenova et al. \cite{Semenova2010} have recently observed an increased concentration of C, N and O at grain boundaries in CP-nTi by atom probe 
tomography, confirming that segregation of light elements is possible at $\alpha$-Ti grain boundaries. However, one could not derive clear 
conclusions about their influence on materials properties, since many factors are acting together. Does 
the segregation correspond to an equilibrium 
state? Does a co-segregation of elements take place? 
Is the formation of clusters at grain boundaries favourable? How strong is the interaction of 
impurities with the GB? Answering these questions solely with experimental methods is very hard or even impossible and requires the use of 
computer simulation \cite{Lymperakis2009}.  Recently, studies of segregation from the first principles have 
become more common 
\cite{Krasko1991,Geng1999,Geng2001,Astala2002,Janisch2003,
Schweinfest2004,Wachowicz2008,Trelewicz2009,Sato2009,Liu2009,Du2011,Abbasi2011,Zhou2011,Zhang2011,Uesugi2012,Sawada2012},
however, in the case of hcp metals such activity is almost absent. In this work we want to fill 
this gap with the example of hcp titanium.  

In summary, a systematic study of segregation in Ti will allows taking further steps towards discovering the most effective mechanism of thermal 
stability in CP-nTi. One may note that quantitative studies of thermal stability of grain boundary structures, including kinetic details of 
mechanisms are most efficiently done by molecular dynamics and Monte Carlo simulations. However, these methods are based on empirical 
interatomic potentials, and therefore a benchmarking with first principles calculations (which are currently not available) is also in this case 
decisive \cite{Lymperakis2009}.  The basis for such calculations is the determination of segregation formation energies, which
are needed for impurity atom to diffuse from a bulk site to a GB site. 

Thus, in this paper we investigate interaction of C, N and O impurities with $\Sigma 7[0001](12\bar{3}0)$ GB in $\alpha$-Ti
from first principles. The results include segregation energies of impurities 
for different positions near and at GB plane. The influence of impurities on 
the grain boundary is accounted by
the relaxation of GB atomic structure. We involve analysis of the electronic structure 
to explain several features of interaction between grain boundary and impurities.

\section{Details of calculation}
\label{details}
Calculations of the full energies and optimized geometries were performed 
in the framework of the density-functional theory (DFT) \cite{Hohenberg1964,Kohn1965}
within the generalized gradient approximation (GGA) 
using the Perdew-Burke-Wang \cite{Perdew1996}  functional and 
projected augmented wave (PAW) method \cite{Blochl1994} (ABINIT \cite{Gonze2002}). 
We considered the following valence
electronic states: $3s,3p,4s,3d$ for Ti and $2s,2p$ for C, N and O. 
To take off any restrictions during relaxation, 
we set number of point group symmetry operations to one in all cases. 
The calculations were performed under three-dimensional periodic boundary conditions. 
An orthorhombic supercell containing 56 Ti atoms was 
constructed using calculated in our previous work \cite{Aksyonov2012} lattice constants of hcp Ti.
The volume of cell was subsequently relaxed to
account the influence of cell sizes and refine the values of lattice constants.
The commensurate supercell containing 54 Ti atoms was employed as a starting model for
the $\Sigma 7[0001](12\bar{3}0)$  GB \cite{Hammerschmidt2005,Sato2005} (Fig. \ref{fig:1}). 
In order to find the optimal geometry of GB, 
the supercell was relaxed with fixed dimensions (as for bulk Ti) in the plane of grain boundary.

The sizes of supercells and lengths $d_{\rm{Ti-Ti}}$
between Ti atoms in
$[10\bar{1}0]$ direction (coincide with $a$ lattice constant in the case of bulk Ti) 
are listed in Table \ref{tab:0}. 
The value of $d_{\rm{Ti-Ti}}$ for bulk Ti supercell is in agreement with experimental lattice constant. In the case of supercell with GB
the variation of $d_{\rm{Ti-Ti}}$ is caused by the influence of grain boundaries and related to the finite sizes of cell.
However, recent studies confirm that selected cell sizes are enough for the description 
of grain boundaries structure
in hexagonal materials \cite{Lane2011} and GB segregation energies \cite{Sawada2012}.

To study the interaction of impurities with the GB, we placed atoms at different positions in optimised supercell
and relax it according to all internal coordinates and size of supercell in the direction perpendicular to the grain boundary plane.
The calculations were performed by using an energy cut-off of 540 eV 
for the plane-wave basis set. 
The Brillouin-zone (BZ) integrals were approximated using the special 
k-point sampling of Monkhorst and Pack \cite{Monkhorst1976}  with 2 $\times$ 1 $\times$ 4 grid.
We have used the Methfessel-Paxton \cite{Methfessel1989}
smearing for Brillouin-zone integration with a smearing width
of 0.027 eV. Such relatively small value does not influence the results but ensures
faster convergence.
The structural optimization was performed until the forces acting on each atom became less than 25 meV/\AA . 
The computational setup ensures that differences in the
segregation energies of interstitials and formation energies of
Ti grain boundaries are converged to within 25 meV and
0.05~J/m$^2$, respectively.
The method of calculation and Ti-C PAW potentials  were checked by computation of lattice and elastic constants of 
several Ti and Ti-C phases in our previous study \cite{Aksyonov2012}.

\begin{table}
\caption{\label{tab:0} Specific excess energy $\gamma$ (J/m$^2$) of grain boundaries 
before and after relaxation, sizes a, b and c (\AA) of considered supercells, and distance $d_{\rm{Ti-Ti}}$ (\AA) between Ti atoms in $[10\bar{1}0]$ direction.} 
\begin{tabular}{lccccccr}
\hline
Cell              & $\gamma$ & a    &   b   &   c     &   $d_{\rm{Ti-Ti}}$ \\
\hline
Bulk$_{relax}$      & -              & 7.77 & 27.03 & 4.62    & 2.95        \\
GB$_{ideal}$       & 1.8             & 7.77 & 27.03 & 4.62    & 2.95        \\
GB$_{relax}$      & 0.73             & 7.77 & 25.97 & 4.62    & 2.75-3.25    \\
\hline

\hline
\end{tabular}
\end{table}

\section{Results and Discussion}
\label{results}
\subsection{Structures and energies of the pure GB}
\label{st}
There are periodic and free boundary conditions (PBC and FBC) which are mainly used for grain boundary modelling.
Free boundary conditions allow to examine one grain boundary in simulation cell 
at the cost of adding two open surfaces. This is useful for the cells with two 
non-equivalent grain boundaries. 
However, there is significant influence 
of open surfaces on atomic structure of supercell. 
The attempt of reduction of such influence by introduction of fixed layers 
cause additional restrictions on the relative moving of adjacent grains during relaxation. 
In the present work, despite the existence of two non-equivalent grain boundaries in simulation cell,
to avoid difficulties related to open surfaces, we use periodic boundary conditions. This is
reasonable due to the specific atomic configuration of cell with one fully coherent 
grain boundary obtained after
relaxation.

As initial structure for grain boundary the ideal symmetric coincidence-site lattice (CSL)
grain boundary $\Sigma 7[0001](12\bar{3}0)$ with $\varphi_{CSL}$=21.8$^\circ$ is used. This boundary
perfectly agrees
with experimentally observed GBs in $\alpha$-Ti  \cite{Wang1996}. 
The procedure of  construction of this GB is 
described elsewhere \cite{Hammerschmidt2005}. 
After the mathematical construction of supercell there are pairs of atoms that lie too close to
each other. Hence, according to \cite{Hammerschmidt2005} each pair was
replaced by one atom.  

\begin{figure*}
\center
\includegraphics[width=15cm]{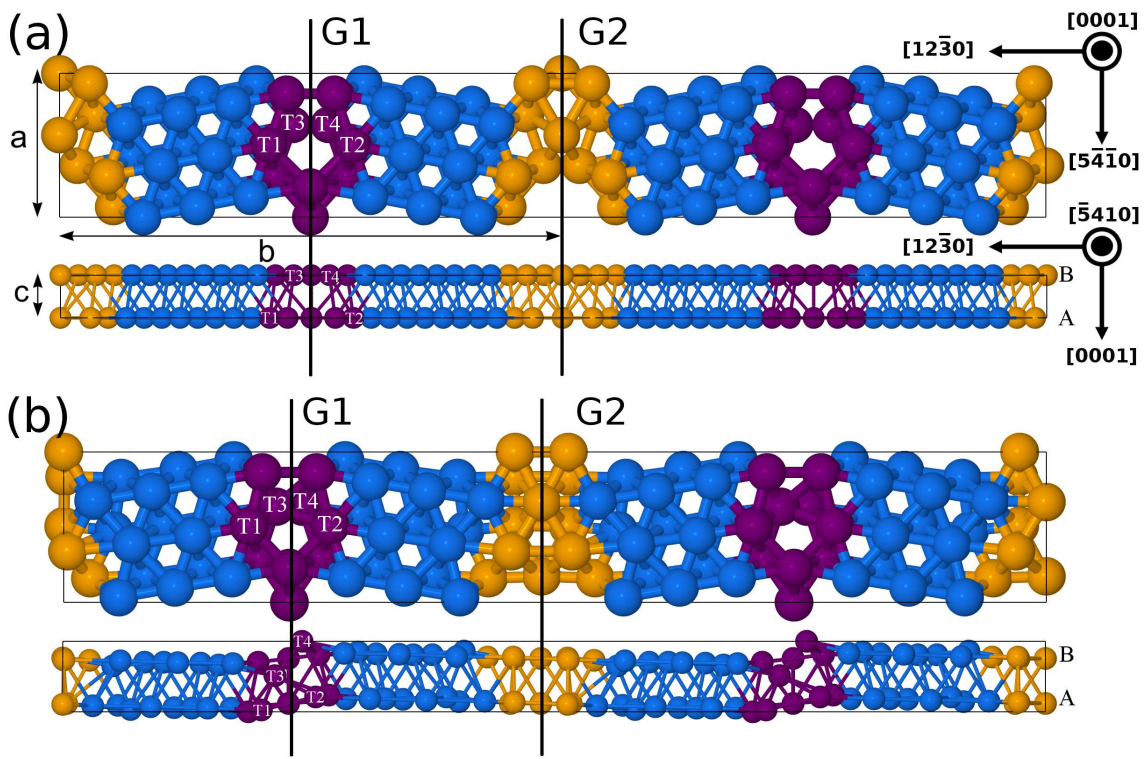}
\caption{\label{fig:1} (a) Structure of the supercell with $\Sigma 7$ GBs after mathematical construction.
(b) Structure of the supercell with $\Sigma 7$ GB after the full relaxation. The shift of grains is clearly visible. To illustrate the structure of the second
GB, two simulation cells are shown. Atoms at GB and bulk regions are painted in different colours. 
The solid lines G1 and G2 denote grain boundary planes.}
\end{figure*}

Two simulation cells are shown in Fig. \ref{fig:1}(a) to indicate the structure of both grain boundaries. 
Ti atoms in the bulk regions and at GBs are shown 
with different colours to guide the eye.
The atomic configuration at G1 and G2 grain boundaries  (see Fig. \ref{fig:1}(a)) in both layers compose of pentagons,
which is in consistent with minimum binding energy structure obtained in \cite{Hammerschmidt2005}.
However, there are differences between structures of G1 and G2 in the bottom layer A (see Fig. \ref{fig:1}(a)). 
The pentagon in this layer  
at G2 is stretched in $[12\bar{3}0]$ direction relative to pentagon at G1 grain boundary. 
Moreover, the upper pentagon at G2 is shifted relative to the bottom pentagon in $[5\bar{4}\bar{1}0]$ direction compared to G1 structure. 
These differences arise from the fact that the coincide site lattice 
exists only in layer B for the given rotation angle.

Two simulation cells after relaxation are shown in Fig. \ref{fig:1}(b). 
The main difference from non-relaxed structure is
that grains are shifted relative to each other along the $[0001]$ direction. 
The interesting feature is that shift
occurs only at G1  plane and is absent at G2. 
The values of shifts $\Delta r_1$ in $[5\bar{4}\bar{1}0]$ direction 
and $\Delta r_2$  in  $[0001]$ direction are averaged by 
several atoms in the bulk region of grains and calculated relative to the initial state. 
The shift is almost absent in $[5\bar{4}\bar{1}0]$
direction for the pure relaxed GB, while the value of $\Delta r_2$ is 0.45 {\AA}.

The atomic structure of G1 and G2 grain boundaries was also significantly changed after relaxation. 
In the case of G1 grain boundary, T3 and T4 atoms
are shifted in opposite directions breaking the symmetry of pentagon and 
forming more close-packed configuration that is more preferable for titanium.
The reduce of cell size in $[12\bar{3}0]$ direction due to the relaxation is $3.7$ \%  (see. Table \ref{tab:0}).

\begin{figure}[h!]
\includegraphics[width=7.5cm]{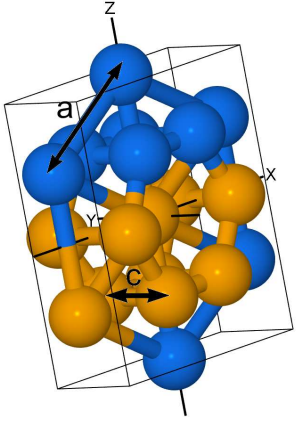}
\caption{\label{fig:12} The structure of G2 grain boundary. The marked atoms correspond to those in Fig. \ref{fig:1}. The additional atoms, 
obtained by periodic replication, complete the conventional cell of the $\omega$-Ti phase in grain boundary. Here a and b corresponds to
the lattice parameters of $\omega$-Ti. The z and y axes are coincide with $[0001]$ and $[12\bar{3}0]$ directions. }
\end{figure}

The atomic structure of G2  is highly symmetrical. 
The visual analysis allows to determine 
$\omega$-Ti phase \cite{Sikka1982rev} at G2. The conventional cell of $\omega$-Ti carved out from G2 is shown 
in Fig. \ref{fig:12}, where marked Ti atoms matches with
Fig. \ref{fig:1}. The orientation relationship is 
$(0001)_\alpha\; \Vert\; (1\bar{2}10)_\omega$ 
and $[12\bar{3}0]_\alpha\; \Vert\; [0001]_\omega$, which is agreed with Silcock \cite{Silcock1958} relationship for planes,
but differs for directions ($[11\bar{2}0]_\alpha\; \Vert\; [0001]_\omega$ for Silcock).
We did not find experimental confirmations of such $[12\bar{3}0]_\alpha\; \Vert\; [0001]_\omega$ relationships
for directions,
but the obtained $\omega-\alpha$ interfaces are quite coherent. The lattice parameters of $\omega$-Ti phase 
at grain boundary are in agreement with that for bulk $\omega$-Ti calculated in \cite{Aksyonov2012}.
The interface period in $[0001]$ direction (4.62 {\AA}) matches very well 
with $a$ parameter of bulk $\omega$-Ti (4.58 {\AA}).
The period in $[5\bar{4}\bar{1}0]$ direction (7.77 {\AA}) matches well with $\sqrt{3}a$ in bulk $\omega$-Ti (7.92 {\AA}).

We study segregation at G1 grain boundary using the PBC, 
because the influence of the coherent G2 grain boundary  
is not larger than that of the open surfaces which emerges due to the FBC. 
Moreover, the use of PBC allows to obtain relative shifts of grains at G1 automatically, as the coherent
G2 grain boundary does not restrict the shift.

Since G1 and G2 grain boundaries are not-equivalent, the following equation allows to calculate
the specific excess energy related to the grain boundaries:
\begin{equation}
\gamma = [E_{GB}(n,a,c)-nE_{sub}(1,a,c)]/S,
\end{equation}
where $E_{GB}(n,a,c)$ is the energy of the cell with GBs and $n$ Ti atoms, 
$E_{sub}(1,a,c)$ is the energy of one atom in the commensurate Ti hcp bulk supercell, and $S=2ac$.
The values of $\gamma$ for the considered grain boundaries before and after relaxation  are 
given in Table \ref{tab:0}. The energy
decreases by more than two times after relaxation.

\subsection{Considered positions of C, N and O atoms}
\begin{figure}
\includegraphics[width=7.5cm]{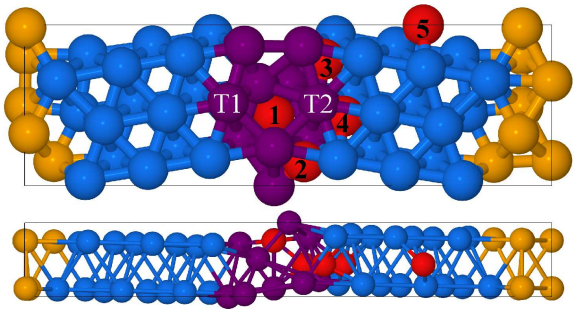}
\caption{\label{fig:3} Atoms 1-5 indicate initial positions used for C, N and O within one relaxed simulation cell. }
\end{figure}

In order to study the interaction of impurities with G1 grain boundary, the same positions for C, N, and O atoms are used.
We have considered five different positions of interstitials within simulation cell: X1 at G1 grain boundary 
(X = C, N, O) and X2-X5 at several distances from the G1 plane. 
The corresponding configurations of all atoms for mentioned positions are also named by X$i$ latter in the text.
Locations of positions are shown within one relaxed simulation cell in Fig. \ref{fig:3},
as the corresponding atomic configurations after relaxation are quite similar for all cases. 
The X2-X5 positions situated in slightly deformed octahedral pores. 
The pore at X1 position has more difficult topology than octahedral one. 

We made additional full relaxation to take into account the possible influence of impurities on 
the grain boundary structure. 
Table \ref{tab:1} contains distance $d$ from the G1 plane to the impurities after relaxation.
The X5 position at the center of two GBs is considered as 
the reference state of impurity in the bulk of titanium.

One may note that distance between impurity and its periodic image in $[0001]$ direction is quite small  
and some interaction is possible. We have checked that this interaction is only 10 \% of the smallest calculated 
value of segregation energy.

\subsection{Segregation energies and volumes of C, N and O atoms at $\alpha$-Ti grain boundary}
The stability of grain boundaries depends on their formation energies \cite{Weissmuller1992}. 
It is well established that the reduction of 
the GB specific energy due to the segregation of some impurities can significantly 
improve the thermal stability of nanostructure \cite{Chookajorn2012,Koch2008}. 

\begin{figure}
\includegraphics[width=7.5cm]{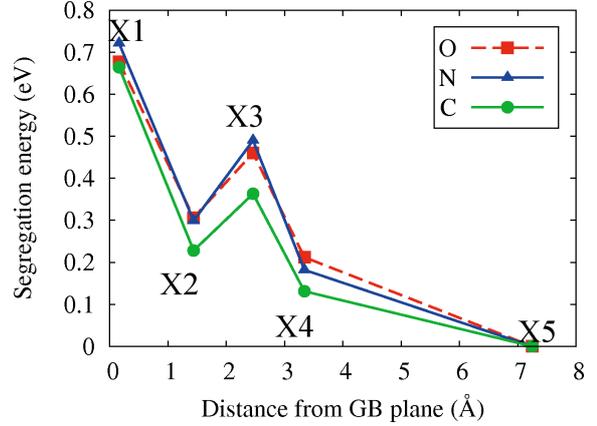}
\caption{\label{fig:4} The dependence of segregation energy E$_{seg}$ from 
the position of impurity.}
\end{figure}

In order to determine possible decrease 
of GB energy due to the impurities, 
the segregation energies and volumes of C, N and O atoms at $\Sigma 7[0001](12\bar{3}0)$ CSL GB in $\alpha$-Ti 
were calculated at 0 K according to the following equations:
\begin{equation}
\begin{aligned}
E_{seg} = E_{\mathrm{Xi}} - E_{\mathrm{X5}} \\
V_{seg} = V_{\mathrm{Xi}} - V_{\mathrm{X5}},
\end{aligned}
\end{equation}
where $E_{\mathrm{Xi}}$ and $V_{\mathrm{Xi}}$ is the energy and volume of the simulation cell with atom X at i position, 
$E_{\mathrm{X5}}$ and $V_{\mathrm{X5}}$ is the energy and volume of the same simulation cell 
with atom X occupying a grain interior site.
The segregation energies and volumes are listed in Table \ref{tab:1}. 
The same trends can be derived
for C, N and O.
The segregation of impurities is unfavourable near $\Sigma 7$ GB in all considered positions.
Moreover, the preference of positions is decreasing towards the G1 grain boundary plane (see Fig. \ref{fig:4}). 
The segregation energies for the same positions almost do not depend on the type of impurity, indicating 
that C, N and O interact with  $\Sigma 7$ GB identically in $\alpha$-Ti.
The increase of segregation energy is accompanied by the increase of segregation volume.
The increase of the volume of the cells is related to the expansion of the G1 grain boundary. 
In particular, the distances between T1 and T2 atoms 
is 4.23 {\AA} for C5, 4.32 {\AA} for C3 and 4.53 {\AA} for C1 configurations. 

\begin{table*}
\center
\caption{\label{tab:1} Segregation energy $E_{seg}$ (eV) and its separation into mechanical E$_{m}$ and chemical E$_{ch}$ contributions.
Geometrical data provided include lateral shifts of grains $\Delta r_1$ in $[5\bar{4}\bar{1}0]$ direction 
and $\Delta r_2$ in $[0001]$ direction (\AA), 
distance $d$ (\AA) between impurity and G1 grain boundary plane, and segregation volume $V_{seg}$ (\AA$^3$) for all considered configurations. The C5, N5 and O5 configurations have the same parameters and denoted as X5.}
\begin{tabular}{lccccccccccccr}
\hline
 & C1 & N1 &  O1 &  C2 &  N2 &  O2 &  C3 &  N3 &  O3 &  C4 &  N4 &  O4 &  X5\\
 \hline
$\Delta r_1$& 0.14&0.16&0.18&--0.08&--0.07& --0.08 & 0.08 & 0.07& 0.08 & 0.06 & 0.07 & 0.06 & 0.04 \\ 
$\Delta r_2$& 0.41&0.49&0.50 & 0.34 & 0.33 & 0.37 & 0.53 & 0.50 & 0.42 & 0.38 & 0.48 & 0.42 & 0.34 \\ 
$d$& 0.16 & 0.35 & 0.35 & 1.44 & 1.44 & 1.44 & 2.46 & 2.49 & 2.51 & 3.34 & 3.32 & 3.33 & 7.25 \\ 
$E_{seg}$&0.66&0.72 & 0.68 & 0.23 & 0.30 & 0.31 & 0.36 & 0.49 & 0.46 & 0.13 & 0.18 & 0.21 & 0.00 \\ 
$E_{m}$& 0.96 & 0.35 & 0.27 & 0.12 & 0.13 & 0.12 & 0.19 & 0.24 & 0.21 & 0.04 & 0.04 & 0.06 & 0.00\\ 
$E_{ch}$& --0.29&0.37 & 0.40 & 0.11 & 0.17 & 0.19 & 0.17 & 0.25 & 0.25 & 0.09 & 0.15 & 0.16 & 0.00\\ 
$V_{seg}$& 8.03 & 4.10 & 3.26 & 2.92 & 4.02 & 3.09 & 0.90 & 3.78 & 2.67 & 0.13 & 1.28 & 1.51 & 0.00\\

\hline
\end{tabular}
\end{table*}

The shifts of grains to each other are listed in Table \ref{tab:1}.
The values of shifts $\Delta r_1$ in the $[5\bar{4}\bar{1}0]$ direction
are quite small (less than 5 \% of titanium lattice constant)
and do not deserve much attention.
However, grains have noticeable additional shifts in $[0001]$ 
direction at G1 plane due to the presence of impurities. The maximum shift is observed for X3 configuration,
which is up to 30 \% greater than $\Delta r_2$ in pure GB.
The dependence of shifts from the position of impurity is rather complex, however in most cases the bigger value
of shift corresponds to the bigger value of $E_{seg}$.

To gain more physical insight on the similarity of segregation energies for different elements,
we separated $E_{seg}$ into mechanical and chemical contributions according to Geng et al. \cite{Geng1999} as follows.
The mechanical contribution $E_{m}$ was determined as the difference between the energies of the structures, 
where the impurity has been removed without subsequent relaxation of the host lattice.
In other words, $E_{m}$ related to the Ti-Ti interaction induced by the impurity.
The chemical contribution
 \begin{equation}
 E_{ch} = E_{seg} - E_{m} 
\end{equation} 
describes direct interaction between impurity and Ti atoms.
The calculated values of $E_{m}$ and $E_{ch}$ are listed in Table \ref{tab:1}.
For all configuration besides C1 one can see the following regularity: (i) The chemical contribution dominates
under $E_{m}$ and correlates with $E_{seg}$; (ii) Values of both contributions for different elements at the same positions are very close;
(iii) The positive values of $E_{m}$ for X2-X4 positions are related to the changes of Ti arrangement at GB and additional
deformation of octahedral pores near the G1 grain boundary; 
(iv) The positive values of $E_{ch}$ connected also with deformation of octahedral pores and depletion of charge density 
near the grain boundary.
In general, such elements as C, N and O  behave in a similar way.
The segregation energy at C1 position with negative chemical contribution is of particular interest and will be discussed below. 

\begin{figure}
\includegraphics[width=7.5cm]{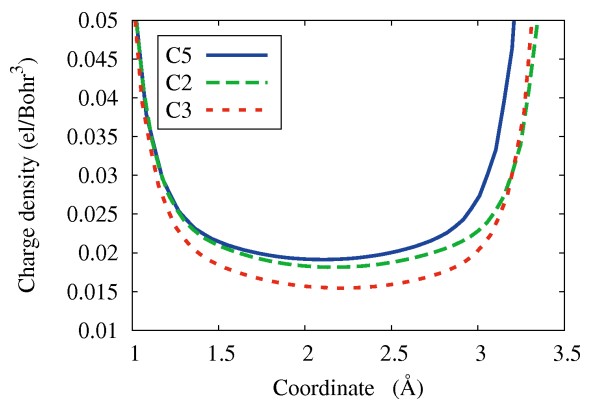}
\caption{\label{fig:5} Line profile of charge density between T1 (0 {\AA}) and T2 (4.22 {\AA}) atoms for the
case of carbon segregation at C5, C2, and C3 positions. The profile for C4 is just below C5 and not shown. The C1 profile is much higher than C5 profile due to the vicinity of carbon atom and also not shown.}
\end{figure}

To explain the positive values of segregation energies, we have studied electronic structure and distribution of charge density. 
The line profile of charge density between T1 and T2 atoms
is shown in Fig. \ref{fig:5} only for the case of carbon. The further discussion will be made for X impurity 
because all three elements demonstrate the same behaviour. 
Carbon, nitrogen and oxygen tends to form covalent $spd$-bonds with Ti atoms accompanied with the
redistribution of charge density: accumulation between Ti-X atoms and depletion between neighbourhood Ti-Ti atoms.
For X2, X3 and in less manner for X4 such redistribution affects the GB region. Depletion of charge density between
T1 and T2 atoms results in a weakening of Ti bonds at G1 GB, increase of their length and increase 
of G1 grain boundary width.
The increase of GB width and volume results in a higher segregation volumes and energies.
The density between T1 and T2 is the highest for X1 position due to the vicinity of impurity, but the atomic size effect leads to the biggest width of GB and the highest segregation energy compared to other positions.

Kwasniak et al. recently obtained from first principles that C, N and O decrease formation energies of
stacking faults in $\alpha$-Ti. This means that segregation of these elements is energetically
favourable at stacking faults. However, authors does not explain the physical reasons for
such behaviour. 
Stacking faults has smaller formation energies than high angle GBs.
Trelewicz et al. \cite{Trelewicz2009} within the framework of statistical thermodynamics
showed that in binary polycrystalline system the segregation energy
decreases (increases for choice of sign of $E_{seg}$ in \cite{Trelewicz2009}) with the reduction of GB formation energy. Hence, the segregation of C, N and O can be possible on low-energy interfaces in $\alpha$-Ti, 
but hardly feasible at high-angle grain boundaries with large formation energies.

\subsubsection{The features of grain boundary segregation}
The segregation energies for X2 and X3 configurations does not fit into the general picture.  
The impurity at X3 position, being slightly far (by 1 \AA) from GB plane than at X2 position,
has higher segregation energy.
The impurity at X3 position attracts T4 titanium atom (see Fig. \ref{fig:1}) reducing T3-T4 spacing. Grain boundary tends to conserve T3-T4
spacing causing additional shift of grains in [0001] direction (see Table \ref{tab:1}). 
In turn, impurity at X2 position
does not change arrangement of Ti atoms at GB and preserve the same shift as in reference X5 position. 
In consequence, additional disturbance of grain boundary structure by impurity at X3 position results in higher segregation energy, 
than by that at X2 position.

\begin{figure}
\includegraphics[width=7.5cm]{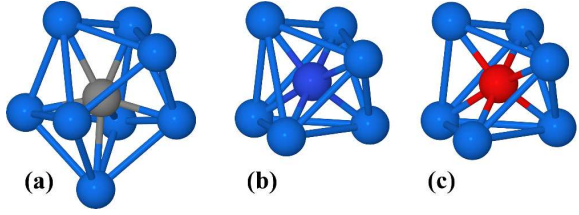}
\caption{\label{fig:71} Local coordination of interstitial
sites at GB for C1 (a), N1 (b) and O1 (c). There are eight Ti neighbours in the case of carbon and 
only six in the case of nitrogen and oxygen. }
\end{figure}

Nevertheless segregation of carbon at C1 position (directly at GB) is unfavourable, the negative chemical contribution 
in segregation energy is observed (see Table \ref{tab:1}).  
It means that the work needed to remove the carbon while not permitting the Ti 
atoms to relax from C1 position is larger by 0.29 eV than that from C5 position.
The comparison of C1 with N1 and O1 shows that in the case of C1 the Ti surrounding coordination is more complex and consists of
eight Ti neighbours. The subtractions of atoms from grain boundaries are show in Fig. \ref{fig:71}. The local coordinations of N1 and O1 
positions are distorted octahedral sites with six Ti neighbours. 
The restructuring of interstitial site in the case of carbon may be related to its higher valency and chemical capacity to
form bonds comparing to N and O. 
However, due to the significant rearrangement of grain boundary structure, the loss in $E_{m}$ overcomes chemical gain. 

\begin{figure}
\includegraphics[width=7.5cm]{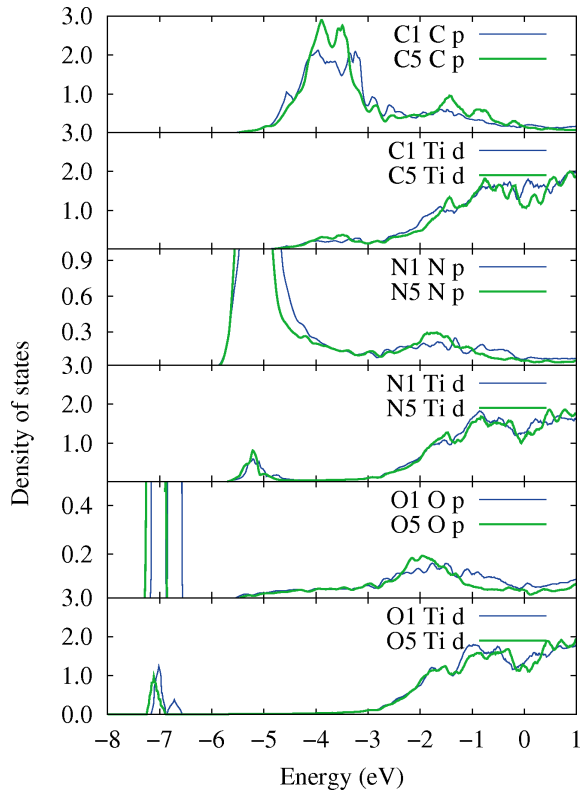}
\caption{\label{fig:8} Site-projected partial density of states of impurities and surrounding titanium atoms for X1 and X5 positions.
The Fermi level lies at the energy zero.}
\end{figure}

To explain the reasons for the negative chemical contribution only in the case of carbon, 
we calculated site-projected partial density of states (PDOS). 
The PDOS of the interstitials ($p$-orbitals) and surrounding Ti atoms ($d$-orbitals) 
for X1 and X5 configurations are shown in Fig. \ref{fig:8}.  
The PDOS of titanium for X1 configuration was obtained by averaging over the Ti atoms shown on 
Fig. \ref{fig:71} and for X5 configuration over the Ti atoms of octahedral coordination around the impurity. 
The most noticeable differences of PDOS between X1 and X5 positions are observed for carbon. 
For the majority of cases one may note the following tendency for PDOS 
which is consistent with the loss of energy for the case of 
impurity at X1 grain boundary site: 
(i) The shift of X1 PDOS to the higher energies relative to X5; 
(ii) Decrease of density of states  at several energies for X1 case.
However, there is qualitative difference for the $p$ states of carbon comparing to oxygen and nitrogen. 
In the case of carbon
the center of gravity of $p$ states at X1 site is shifted to lower energies ($-0.067$ eV), 
 while centres of gravity of N and O $p$ bands at X1 are shifted
to the higher energies ($+0.124$ eV and $+0.113$ eV respectively). 
This can be connected with the negative chemical contribution in the case of C1 configuration.

The smaller values of $E_{m}$ for N and O is in consistent with the lower values of $V_{seg}$
which means that N and O atoms
require less space at the grain boundary than carbon.

Summing up the results, it should be mentioned that the used method for $E_{seg}$ calculation has one drawback related
to the optimisation of GB structure. The drawback is linked to the fact that it is hardly possible to obtain 
by first principles the shift of grains which corresponds to the global minimum of total energy.
There is a possibility of relaxation of compared configurations to different local minima of energy. 
However, the obtained 
self-consistent picture of the interaction between impurity and GB
allows us to claim that qualitative results will remain unchanged for the true ground states.
Moreover, the quantitatively correct results have little sense for the considered case, since experimentally measured segregation energies are always statistically averaged over a large number of different sites at general grain boundaries.
\subsubsection{Comparison with experimental data}
A few words should be said about agreement of our results with experimental observation of increased concentration 
of C, N and O at GBs in Ti by Semenova et al. \cite{Semenova2010}.
Primarily, we do not claim that equilibrium segregation of these elements is principally impossible in $\alpha$-Ti, 
since we considered only one type of grain boundary. 
However, there are also two interpretations for the experimental data can be done as authors do not give any.
Semenova et al. provide atom probe tomography composition profiles of interstitial impurities across the grain boundary 
after annealing at 623 K (see Fig. 4 in \cite{Semenova2010}).
On the one hand, the asymmetry and significant width of the observed concentration peaks are usually related to
the non-equilibrium segregation
due to the grain boundary migration \cite{Lejcek2010}. On the other hand, the correlation between peaks for carbon and oxygen
suggests that C-O co-segregation had been observed. Hence, there is an intriguing possibility for 
grain boundary C-O co-segregation in $\alpha$-Ti without its independent segregation.

Speaking about the possible mechanisms of thermal stabilisation in nanocrystalline 
pure titanium, the results of the present work are in favour
of grain boundary pinning by bulk precipitates formed from solid solution~\cite{Aksyonov2012}.

\section{Conclusions}
\label{conc}
In summary, we have investigated interaction of C, N and O atoms with 
high-angle $\Sigma 7[0001](12\bar{3}0)$ grain boundary (GB) in $\alpha$-Ti through the
\textit{ab initio} PAW potential calculations. 
The specific simulation cell with two non-identical GBs 
have allowed to use the full periodic boundary conditions due to the coherency of the second GB.
The first GB has the structure of unsymmetrical tilt boundary with lateral shifts, 
while the second GB replicates the structure of $\omega$-Ti  phase with
$(0001)_\alpha \; \Vert \;(1\bar{2}10)_\omega$ 
and $[12\bar{3}0]_\alpha\; \Vert\; [0001]_\omega$ orientation relationship. 

The segregation
energies were calculated for different positions at different distances from GB plane. It was obtained that all considered elements
has similar behaviour and prefer to be in octahedral site of bulk Ti. The maximum energy loss is observed for the positions at GB interface.
The analysis of geometry and electronic structure showed that the 
increase of energy is due to the redistribution of electronic density and reduce of the bonding strength 
within the grain boundary area. 

Therefore, we believe that grain boundary pinning by small Ti-C particles is the main mechanism of thermal stability in
nanostructured titanium \cite{Aksyonov2012}. 
We are aware, however, that in a real system more complex processes such as grain boundary co-segregation and 
precipitation could occur, making further theoretical and experimental investigations desirable.

\section{Acknowledgements}
The research led to these results received funding from
the Federal Target Program under Grant Agreement No. 2.2437.2011 and No. 14.A18.21.0078.
We are grateful to the "SKIF-polytech" research group 
for providing computational resources on the "SKIF Cyberia" cluster at the Tomsk State University.






\bibliographystyle{model1a-num-names}
\bibliography{library}






\end{document}